\documentclass[12pt]{article}

\usepackage[margin=0.55in]{geometry}

\usepackage{amsmath}
\usepackage{amsfonts}
\usepackage{amsthm}
\usepackage{array}
\usepackage{multirow}
\usepackage{graphicx}
\usepackage{caption}
\usepackage{subcaption}

\DeclareMathOperator{\diver}{div}

\newcommand{\totalDiff}[1]{\frac{\mathrm{d}}{\mathrm{d}{#1}}}
\newcommand{\parder}[2]{\frac{\partial #1}{\partial #2}}

\newcommand{\dir}[1]{\partial_{ {#1}}}

\newcommand{\LieDerivative}[2]{\mathcal{L}_{#1}\left(#2\right)}

\newcommand{\LieAlgebra}[1]{\mathfrak{#1}}

\newcommand{\bvec}[1]{\mathbf{#1}}
\newcommand{\entr}{s}
\newcommand{\temp}{T}
\newcommand{\dens}{\rho}
\newcommand{\press}{p}
\newcommand{\energy}{\epsilon}

\newcommand{\systemEk}[1]{\mathcal{E}_{#1}} 
\newcommand{\stress}{\sigma}

\newtheorem{theorem}{Theorem}

\title{Symmetries and Differential Invariants for Viscid
Flows on a Curve}
\author{Anna Duyunova,
	\\ Institute of Control Sciences of RAS,
	\\ duyunova\_anna@mail.ru,\\
	Valentin Lychagin,
	\\ Institute of Control Sciences of RAS, University of Troms\o,\\
	valentin.lychagin@uit.no,\\
	Sergey Tychkov,
	\\ Institute of Control Sciences of RAS,\\ sergey.lab06@yandex.ru
}
\date{}

\begin{document}
\maketitle 	
\abstract{In this paper, flows of a viscid fluids on curves
are considered. Symmetry algebras and the corresponding
fields of differential invariants are found. We study their
dependence on thermodynamic states of media, and provide
classification of thermodynamic states.}
	
\section{Introduction}

Consider flow of an viscid medium on an oriented Riemannian
manifold $(M,g)$ in the field of a constant gravitational
field. Motion of viscous media are described by the PDE
system consisting of the Navier-Stokes equation, the laws
of mass and energy conservation (see \cite{Batchelor2000},
\cite{DLTNS} for details):

\begin{equation}\label{eq:E}
\left\lbrace
\begin{aligned}
&\dens(\bvec{u}_t+\nabla_{\bvec{u}}\bvec{u})-\diver\stress-\bvec{g}\dens=0,\\
&\parder{(\dens\,\Omega_g)}{t}+\LieDerivative{\bvec{u}}{\dens\,\Omega_g} = 0,\\
&\dens\temp\left(\entr_t +\nabla_{\bvec{u}}s\right)-\Phi+k(\Delta_g\temp) =0,
\end{aligned}
\right.
\end{equation}
where the vector field $\bvec{u}$ is the flow velocity,
$\press$, $\dens$, $\entr$, $\temp$ are the pressure,
density, specific entropy, temperature of the fluid
respectively, $k$ is the thermal conductivity, which is
supposed to be constant, and $\bvec{g}$ is the gravitational
acceleration. The stress tensor $\stress$ depends on two
viscousites, which are alse considered constant.

In this paper, we consider the case, when $M$ is a
naturally-parameterized curve in the three-dimensional
Euclidean space
\[
M=\{x=f(a),\,y=g(a),\,z=h(a)\} 
\]
In this case, vector $\bvec{g}$ is the restriction
of the vector field $(0,0,\mathrm{g})$ on $M$, i.e.,
\[
\bvec{g}=\mathrm{g}h^{\prime}\partial_a.
\]

First of all, we should note that two additional relations
involving thermodynamic quantites are needed to complete the
system~\eqref{eq:E}. To obtain them, we apply the method
described in the paper \cite{DLTwisla} in detail. The
general idea behind this method is representation of
thermodynamic states with Legendrian, or Lagrangian,
manifolds in a contact, or symplectic, space
correspondingly.

So, by the Navier-Stokes system $\systemEk{}$ we mean the
equations \eqref{eq:E} together with two equations of the
thermodynamic state
\begin{equation}\label{eq:Therm}
L= \{\, F(\press,\dens,\entr,\temp)=0,\,
G(\press,\dens,\entr,\temp)=0 \,\}
\end{equation}
that meet the condition 
\[
[F,G]=0 \quad\mathrm{mod}\quad\{F=0,\,G=0\},
\]
where $[F,G]$ is the Poisson bracket with respect to the
symplectic form 
\[
\Omega=d\entr\wedge d\temp+{\dens^{-2}}d\dens\wedge d\press.
\]
Moreover, the restriction of the quadratic differential form
\[
\kappa= d(\temp^{-1})\cdot d\energy -
\dens^{-2} d(\press \temp^{-1}) \cdot d\dens
\]
to the manifold of thermodynamic state is negative definite,
here $\energy$ is the specific internal energy. 

The paper is organized as follows.

In Section \ref{sec:symmetries} we study symmetry Lie
algebras of the Navier-Stokes system $\systemEk{}$  and
their dependence on the form of the function $h(a)$. There
are six different forms, besides the general case, of the
function $h$ that correspond to different symmetry algebras.

In Section \ref{sec:therm} we consider the case when the
thermodynamic state admits a one-dimensional symmetry
algebra and find the corresponding Lie algebras. For such
thermodynamic states, we find an explicit form of Lagrangian
surface in terms of two equations on the thermodynamic
quantities $\press$, $\temp$, $\dens$ and $\entr$.

In Section \ref{sec:invs} we recall the notion of
differential invariants and introduce Navier-Stokes and
kinematic invariants. For these types we find the field of
differential invariants.

A space curve can be represented as a lift of a plane curve.
Connection between the function $h$ and a way of lifting
curve was discussed in \cite{DLTcurve}.

Most of the computations in this paper were done in Maple
with the Differential Geometry package by I. Anderson and
his team \cite{Anderson2016}.

\section{Symmetry Lie algebra}\label{sec:symmetries}

Using the standard techniques for calculating of symmetries
we find dependence of symmetry algebra of system
$\systemEk{}$ on the function $h(a)$ (see the Maple files
\textit{http://d-omega.org}).

To this end, we consider a Lie algebra $\LieAlgebra{g}$ of
point symmetries of the system \eqref{eq:E}.

Let $\vartheta\colon\LieAlgebra{g} \rightarrow \LieAlgebra{h}$
be the following Lie algebras homomorphism
\[
\vartheta\colon X\mapsto
X(\dens)\dir{\dens} + X(\entr)\dir{\entr} +
X(\press)\dir{\press} + X(\temp)\dir{\temp},
\]
where $\LieAlgebra{h}$ is a Lie algebra generated by vector
fields that act on the thermodynamic valuables $\press$,
$\dens$, $\entr$ and $\temp$.

The kernel of the homomorphism $\vartheta$ is an ideal
$\LieAlgebra{g_{m}}\subset \LieAlgebra{g}$, and we call the
elements of $\LieAlgebra{g_{m}}$ {\it geometric
symmetries}.

Let $\LieAlgebra{h_t}$ be a such Lie subalgebra of the
algebra $\LieAlgebra{h}$ that preserves thermodynamic state
\eqref{eq:Therm}.

Then the following theorem is true (see for details
\cite{DLTwisla}).
\begin{theorem}
	A Lie algebra $\LieAlgebra{g_{sym}}$ of symmetries of
	the Navier-Stokes system $\systemEk{}$ coincides with 
	\[
	\vartheta^{-1}(\LieAlgebra{h_{t}}).
	\]
\end{theorem}


First of all, consider the general case, when $h(a)$ is an
arbitrary function. Then the Lie algebra $\LieAlgebra{g^0}$
of point symmetries of the system \eqref{eq:E} is generated
by the vector fields 
\begin{equation*} 
X_1 = \dir{ t}, \qquad  
X_{2} = \dir{ \press} , \qquad  
X_{3} =  \dir{ \entr}  .
\end{equation*}

The pure thermodynamic part $\LieAlgebra{h^0}$ of the system
symmetry algebra in this case is generated by
\begin{equation*} 
Y_1 = \dir{ \press}, \qquad  
Y_{2} = \dir{ \entr}  .
\end{equation*}

The PDE system $\systemEk{}$ has the
smallest Lie algebra of point symmetries $
\vartheta^{-1}(\LieAlgebra{h_{t}^0})$, when the function
$h(a)$ is arbitrary.

The special cases of the function $h(a)$ are listed below.

\medskip
\textbf{1.} $h(a)= const$ 

When the symmetry Lie algebra $\LieAlgebra{g^1}$ of sthe
system \eqref{eq:E} is generated by $X_1,X_2,X_3$ and by the
following vector fields
\[
\begin{aligned} 
&X_4 = \dir{ a}, \qquad \qquad \,\,\,\,
X_6 = t\,\dir{ t}+a\,\dir{a}-\press\,\dir{\press} -\dens\,\dir{ \dens}, \\
&X_5 = t\,\dir{ a}+\dir{ u}, \qquad
X_7 = a\,\dir{a} + u\,\dir{ u}-2\dens\,\dir{\dens}
+2\temp\,\dir{\temp}.
\end{aligned} 
\]

The Lie algebra $\LieAlgebra{g^1}$ is solvable and the
sequence of derived algebras is the following 
\[
\LieAlgebra{g^1} = \left\langle X_1,X_2,\ldots,X_7
\right\rangle \supset
\left\langle X_1,X_2,X_3,X_4,X_5\right\rangle\supset
\left\langle X_4\right\rangle=0.
\]

The pure thermodynamic part $\LieAlgebra{h^1}$ of the
symmetry algebra is generated by the vector fields 
\begin{equation*} 
Y_1=\dir{\press},\qquad  
Y_2=\dir{\entr},\qquad  
Y_3=\press\,\dir{\press}+\dens\,\dir{\dens},\qquad 
Y_4=\dens\,\dir{\dens}-\temp\,\dir{\temp}.
\end{equation*}

Hence, the PDE system $\systemEk{}$ admits a Lie algebra of
point symmetries $\vartheta^{-1}(\LieAlgebra{h_{t}^1})$.

\medskip
\textbf{2.} {$h(a)= \lambda a$,  $\lambda\neq 0$}

In this case the Lie algebra $\LieAlgebra{g^2}$ of point
symmetries of the system \eqref{eq:E} is generated by
$X_1,X_2,X_3$ and by the following vector fields
\begin{align*} 
&X_4 = \dir{a},\qquad\quad\quad\,\,\,\,
X_6=t\,\dir{t}+2a\,\dir{a}+u\,\dir{u}-\press\,\dir{\press}-
3\dens\,\dir{\dens}+2\temp\dir{\temp},\\
&X_5=t\,\dir{a}+\dir{u},\qquad
X_7=t\,\dir{t}+ (\frac{\lambda gt^2}{2}+a)\,\dir{a}+
\lambda gt\,\dir{u}-\press\,\dir{\press}-\dens\,\dir{\dens}.
\end{align*} 

The Lie algebra $\LieAlgebra{g^2}$ is solvable and its
sequence of derived algebras is
\[
\LieAlgebra{g^2}=\left\langle X_1,X_2,\ldots,X_7 \right\rangle
\supset\left\langle X_1,X_2,X_3,X_4,X_5\right\rangle\supset
\left\langle X_4\right\rangle=0.
\]

The pure thermodynamic part $\LieAlgebra{h^2}$ of the
symmetry algebra is generated by the vector fields 
\begin{equation*} 
Y_1=\dir{\press}, \qquad  
Y_2= \dir{ \entr} , \qquad  
Y_3= \press\,\dir{ \press}+\dens\,\dir{ \dens} ,\qquad 
Y_4=\dens\,\dir{ \dens}  - \temp\,\dir{ \temp} .
\end{equation*}

Hence, the PDE system $\systemEk{}$ admits a Lie algebra of
point symmetries $\vartheta^{-1}(\LieAlgebra{h_{t}^2})$.

\medskip
\textbf{3.} $h(a)= \lambda a^2$, $\lambda\neq 0$

In this case the Lie algebra $\LieAlgebra{g^3}$ of point
symmetries of the system \eqref{eq:E} is generated by the
vector fields $X_1,X_2,X_3$ and, if $\lambda<0$, by the
vector fields 
\[
\begin{aligned} 
&X_4 = \sin(\sqrt{2\lambda \mathrm{g}}\,t)\,\dir{a} +
\sqrt{2\lambda \mathrm{g}} \cos(\sqrt{2\lambda
\mathrm{g}}\,t)\,\dir{u},\qquad\\
&X_5 =\cos(\sqrt{2\lambda \mathrm{g}}\,t)\,\dir{a} -
\sqrt{2\lambda \mathrm{g}} \sin(\sqrt{2\lambda
\mathrm{g}}\,t)\,\dir{u},\\
&X_6 = a\, \dir{ a} +u\,\dir{ u}  
 -2\dens\,\dir{ \dens} +
{2  }\temp\dir{ \temp} 
\end{aligned} 
\]
and, if $\lambda>0$, by the vector fields 
\[
\begin{aligned} 
&X_4 = \exp(\sqrt{-2\lambda \mathrm{g}}\,t)\,\dir{a} +  \sqrt{-2\lambda \mathrm{g}} \exp(\sqrt{-2\lambda \mathrm{g}}\,t)\,\dir{ u} , \qquad \\
&X_5 =\exp(-\sqrt{-2\lambda \mathrm{g}}\,t)\,\dir{a} -  \sqrt{-2\lambda \mathrm{g}} \exp(-\sqrt{-2\lambda \mathrm{g}}\,t)\,\dir{ u} , \\
&X_6 = a\, \dir{ a} +u\,\dir{ u}  
-2\dens\,\dir{ \dens} +
{2  }\temp\dir{ \temp} .
\end{aligned}  
\]

The Lie algebra $\LieAlgebra{g^3}$ is solvable and its
sequence of derived algebras is 
\[
\LieAlgebra{g^3} = \left\langle X_1,X_2,\ldots,X_6\right
\rangle\supset
\left\langle X_2,X_3,X_4,X_5\right\rangle=0.
\]

The pure thermodynamic part $\LieAlgebra{h^3}$ of the
symmetry algebra is generated by the vector fields 
\begin{equation*} 
Y_1=\dir{\press},\qquad  
Y_2=\dir{\entr},\qquad  
Y_3=\dens\,\dir{\dens}-\temp\,\dir{\temp}.
\end{equation*}

Hence, the PDE system $\systemEk{}$ admits a Lie algebra of
point symmetries $\vartheta^{-1}(\LieAlgebra{h_{t}^3})$.

\medskip
\textbf{4.} $h(a)= \lambda_1a^{\lambda_2}$, $\lambda_2\neq 0,1,2$ 

The Lie algebra $\LieAlgebra{g^4}$ of point symmetries of
the system \eqref{eq:E} is generated by the vector fields
$X_1,X_2,X_3$ and by the vector field
\[
\begin{aligned}
&X_4 = t\,\dir{ t} -\frac{2a}{\lambda_2-2} \dir{ a} -
\frac{\lambda_2 u }{\lambda_2-2}\dir{ u} - 
 \press\,\dir{ \press}  +
\frac{\lambda_2 + 2  }{\lambda_2-2}\dens\,\dir{ \dens} -
 \frac{2\lambda_2  }{\lambda_2-2}\temp\,\dir{ \temp} .
\end{aligned}
\]

The Lie algebra $\LieAlgebra{g^4}$ is solvable and the
sequence of derived algebras is the following 
\[
\LieAlgebra{g^4} = \left\langle X_1,X_2,X_3,X_4 \right\rangle\supset
\left\langle X_1,X_2,X_3 \right\rangle=0 .
\]

The pure thermodynamic part $\LieAlgebra{h^4}$ of the
symmetry algebra is generated by the vector fields 
\[
Y_1=\dir{\press},\qquad  
Y_2=\dir{\entr},\qquad
Y_3=\press\,\dir{\press}-(\lambda_2+2)\dens\,\dir{\dens}+
2\lambda_2 \temp\,\dir{ \temp}.
\]

Hence, the PDE system $\systemEk{}$ admits a Lie algebra of
point symmetries $\vartheta^{-1}(\LieAlgebra{h_t^4})$.

\medskip
\textbf{5.} $h(a)=\lambda_1e^{\lambda_2a}, \, \lambda_2\neq 0$

In this case, the symmetry Lie algebra $\LieAlgebra{g^5}$ of
the system \eqref{eq:E} is generated by the vector fields
$X_1,X_2,X_3$ and by the vector field
\[
X_4 = t\,\dir{ t}-\frac{2}{\lambda_2}\,\dir{a} - u\,\dir{ u}  - \press\,\dir{ \press}  +
\dens\,\dir{ \dens} -2\temp\dir{ \temp}.
\]

The Lie algebra $\LieAlgebra{g^5}$ is solvable and the
derived algebras are the following 
\[
\LieAlgebra{g^5} = \left\langle X_1,X_2,X_3,X_4 \right\rangle\supset
\left\langle X_1,X_2,X_3 \right\rangle=0 .
\]

The pure thermodynamic part $\LieAlgebra{h^5}$ of the
symmetry algebra is generated by the vector fields 
\[
Y_1=\dir{\press},\qquad  
Y_2=\dir{\entr},\qquad
Y_3=\press\,\dir{\press}-\dens\,\dir{\dens}+
2\temp\,\dir{\temp}.
\]

Hence, the PDE system $\systemEk{}$ admits a Lie algebra of
point symmetries $\vartheta^{-1}(\LieAlgebra{h_{t}^5})$.

\medskip
\textbf{6.} $h(a)= \ln a$

The Lie algebra $\LieAlgebra{g^6}$ of point symmetries of
the system \eqref{eq:E} is generated by the vector fields
$X_1,X_2,X_3$ and by the vector field
\[
X_4= t\,\dir{ t}+ a  \,\dir{a}- \press\,\dir{ \press} -\dens\,\dir{\dens}.
\]

The Lie algebra $\LieAlgebra{g^6}$ is solvable and the
sequence of derived algebras is the following 
\[
\LieAlgebra{g^6} = \left\langle X_1,X_2,X_3,X_4 \right\rangle \supset
\left\langle X_1,X_2 \right\rangle =0.
\]

The pure thermodynamic part $\LieAlgebra{h^6}$ of the
symmetry algebra is generated by the vector fields 
\[
Y_1=\dir{\press},\qquad  
Y_2=\dir{\entr},\qquad  
Y_3=\press\,\dir{\press}+\dens\,\dir{\dens}.
\]

Hence, the PDE system $\systemEk{}$ admits a Lie algebra of
point symmetries $\vartheta^{-1}(\LieAlgebra{h_{t}^6})$.

\section{Thermodynamic states with a one-dimensional symmetry algebra} \label{sec:therm}

In this section we consider the thermodynamic states, or the
Lagrangian surfaces $L$, admitting a one-dimensional
symmetry algebra. The cases, when thermodynamic states admit
a two-dimensional symmetry algebra, can be studied in the
similar way.

Let the thermodynamic state admit a one-dimensional symmetry
algebra. Denote by
\[
Z=\gamma_1Y_1+ \gamma_2Y_2+ \ldots + \gamma_kY_k
\]
a basis vector of this algebra, then the Lagrangian surface
can be found from the solution of PDE
(see \cite{DLTwisla} for details) 
\[
\left\{\begin{aligned}
&\Omega\vert_{L}=0,\\
&(\iota_{Z}{\Omega})\vert_{L}=0.
\end{aligned}\right.
\]

This system in terms of specific energy can be written as
\[
\energy=\energy(\dens,\entr),\quad\temp=\energy_{\entr},
\quad\press=\dens^2\energy_{\dens}.
\]
Solving it, we find thermodynamic state $L$, which must also
satisfy $\kappa\vert_{L}<0$.

Straightforward computations show that, for an arbitrary
function $h(a)$, there are no thermodynamic states that
admit a one-dimensional symmetry algebra.

\medskip
\textbf{1, 2.} $h(a)=const$, $h(a)=\lambda a$

The pure thermodynamic part of the system symmetry algebra
coincides with the thermodynamic part of the 2d
Navier-Stokes case. Thus, the classification of the
thermodynamic states for these two cases can be found in
\cite{DLTwisla}.
\medskip
\textbf{3.} {$h(a)=\lambda a^2$, $\lambda\neq 0$}

Let a basis vector of a one-dimensional symmetry algebra be
\[
\gamma_1\dir{ \press} +\gamma_2\dir{ \entr}  + \gamma_3
( \dens\,\dir{ \dens}  -\temp\,\dir{ \temp} ),
\]
then in the general case expressions for the pressure and
temperature have the form
\[
\press = \frac{\gamma_2}{\gamma_3} F^{\prime} -F -
\frac{\gamma_1}{\gamma_3}(\ln\dens-1),\quad
\temp = \frac{F^{\prime}}{\dens},\quad
F=F\left(\entr + \frac{\gamma_2}{\gamma_3}\ln\dens \right),
\]
where $F$ is an arbitrary function. The condition of
negative definiteness of the differential form $\kappa$
leads to the relations  
\[
F^{\prime}>0, \quad  F^{\prime\prime}>0, \quad
\frac{(\gamma_2F^{\prime} - \gamma_1)F^{\prime\prime}}{\gamma_3}  - F^{\prime 2}>0.
\]

\medskip
\textbf{4.} $h(a)= \lambda_1a^{\lambda_2}$, $\lambda_2\neq 0,1,2$
 
Let a basis vector of a one-dimensional symmetry algebra be
\[
\gamma_1\dir{\press}+\gamma_2\dir{\entr}+
\gamma_3( \press\,\dir{ \press}- (\lambda_2 + 2)\dens
\,\dir{\dens} + 2\lambda_2 \temp\,\dir{\temp}),
\]
then in the general case expressions for the pressure and
temperature have the form
\[
\press = \frac{\dens^{\frac{2-\lambda_2}{\lambda_2+2}}
(\gamma_2F^{\prime}-2\lambda_2\gamma_3F)}{\gamma_3(\lambda_2+2)} - \frac{\gamma_1}{\gamma_3(\lambda_2-2)}, \quad
\temp = \dens^{\frac{-2\lambda_2}{\lambda_2+2}} F^{\prime}, \quad
F= F\left(\entr + \frac{\gamma_2}{\gamma_3(\lambda_2+2)}\ln\dens \right) ,
\]
where $F$ is an arbitrary function. The negative
definiteness of the differential form $\kappa$ leads to the
relations 
\[
F^{\prime}>0, \quad  F^{\prime\prime}>0, \quad
2\lambda_2(\lambda_2-2)F F^{\prime\prime} -4\lambda_2^2F^{\prime2} +
\frac{\gamma_2(\lambda_2+2)F^{\prime}F^{\prime\prime}}{\gamma_3}>0.
\]

\medskip
\textbf{5.} $h(a)=\lambda_1e^{\lambda_2}$

The pure thermodynamic part of the system symmetry algebra
coincides with the symmetry Lie algebra of the Navier-Stokes
system of differential equations on a two dimensional unit
sphere. So, the classification of thermodynamic states can
be found in \cite{DLTwisla}.

\medskip
\textbf{6.} $h(a)= \ln a$

Let a basis vector of a one-dimensional symmetry algebra be 
\[
\gamma_1\dir{\press}+\gamma_2\dir{\entr}+\gamma_3
(\press\,\dir{ \press}+\dens\,\dir{ \dens}),
\]
then in the general case expressions for the pressure and
temperature have the form
\[
\press = \frac{-(\gamma_2 F^{\prime} + C )\dens}{\gamma_3}-
\frac{\gamma_1}{\gamma_3}, \quad
\temp = F^{\prime}, \quad
F=F\left( \entr- \frac{\gamma_2}{\gamma_3}\ln \dens \right). 
\]
The negative definiteness of the differential form $\kappa$
leads to the relations 
\[
F^{\prime}>0, \quad  F^{\prime\prime}>0, \quad
\frac{\gamma_2 F^{\prime} + C}{\gamma_3}<0
\]
when $\entr \in(-\infty,\entr_0]$.


\section{Differential invariants} \label{sec:invs}

As before in \cite{DLTwisla}, we consider two group actions
on the Navier-Stokes system $\systemEk{}$. Specifically, the
prolonged actions of the Lie algebras $\LieAlgebra{g_{m}}$
and $\LieAlgebra{g_{sym}}$.

Recall that a function $J$ on the manifold $\systemEk{k}$ is
a \textit{kinematic differential invariant of order} $\leq
k$ if 
\begin{enumerate}
	\item $J$ is a rational function along fibers of the
	projection $\pi_{k,0}\colon\systemEk{k}\rightarrow
	\systemEk{0}$,
	\item $J$ is invariant with respect to the prolonged
	action of the Lie algebra $\LieAlgebra{g_{m}}$,
	i.e., for all $X\in \LieAlgebra{g_{m}}$,
	\begin{equation} \label{dfinv}
	X^{(k)}(J)=0,
	\end{equation}
\end{enumerate}
where $\systemEk{k}$ is the prolongation of the system
$\systemEk{}$ to $k$-jets, and $X^{(k)}$ is the $k$-th
prolongation of a vector field $X\in \LieAlgebra{g_{m}}$.

Note that fibers of the projection $\systemEk{k}\rightarrow
\systemEk{0}$ are irreducible algebraic manifolds.

A kinematic invariant is \textit{an Navier-Stokes invariant}
if condition \eqref{dfinv} holds for all $X \in
\LieAlgebra{g_{sym}}$.

We say that a point $x_k\in \systemEk{k}$ and the
corresponding orbit $\mathcal{O}(x_k)$
($\LieAlgebra{g_{m}}$- or $\LieAlgebra{g_{sym}}$-orbit) are
\textit{regular}, if there are exactly $m=\mathrm{codim}\,
\mathcal{O}(x_k) $ independent  invariants (kinematic or
Navier-Stokes) in a neighborhood of this orbit. Otherwise,
the point and the corresponding orbit are \textit{singular}.

The Navier-Stokes system together with the symmetry algebras
$\LieAlgebra{g_{m}}$ or $\LieAlgebra{g_{sym}}$ satisfies
the conditions of Lie-Tresse theorem (see \cite{KL}), and,
therefore, the kinematic and Navier-Stokes differential
invariants separate regular $ \LieAlgebra{g_{m}}$ and
$\LieAlgebra{g_{sym}}$ orbits on the Navier-Stokes system
$\systemEk{}$ correspondingly. 

By a $\LieAlgebra{g_{m}}$ or
$\LieAlgebra{g_{sym}}$-invariant differentiation we mean
a total differentiation
\[
A\totalDiff{t}+B\totalDiff{a}
\]
that commutes with prolonged action of algebra
$\LieAlgebra{g_{m}}$  or $\LieAlgebra{g_{sym}}$. Here $A$,
$B$ are rational functions on the prolonged system
$\systemEk{k}$ for some $k\geq 0$.

\subsection{Kinematic invariants}
\begin{theorem} 
	\begin{enumerate}
		\item The field of kinematic invariants is generated
		by first-order basis differential invariants and by
		basis invariant differentiations. This field
		separates regular orbits.
		\item For the general cases of $h(a)$, as well as
		for
		$h(a)=\lambda_1a^{\lambda_2}$,
		$h(a)=\lambda_1e^{\lambda_2a}$ and $h(a)=\ln a$,
		the basis differential invariants are
		\[
		a,\quad u,\quad \dens,\quad\entr,\quad  u_t,\quad
		u_a,\quad\dens_a,\quad \entr_t, \quad\entr_a,
		\]
		and the basis invariant differentiations are
		\[
		\totalDiff{t} ,\quad \totalDiff{a} . 
		\]
		\item For the cases $h(a)=const$, $h(a)=\lambda a$ 
		the basis differential invariants are
		\[
		\dens,\quad\entr,\quad u_a, \quad u_t+u u_a, \quad\dens_a, \quad \entr_a, \quad \entr_t+u \entr_a,
		\]
		and basis invariant differentiations are
		\[
		\totalDiff{t} + u \totalDiff{a} ,\quad  \totalDiff{a}.
		\]	
		\item For the case $h(a)=\lambda a^2$  
		the basis differential invariants are
		\[
		\dens,\quad\entr,\quad u_a, \quad u_t+u u_a-2\lambda g a, 
		\quad\dens_a, \quad \entr_a, \quad \entr_t+u \entr_a ,
		\]
		and basis invariant differentiations are
		\[
		\totalDiff{t}+u\totalDiff{a},\quad\totalDiff{a}.
		\]	
		\item The number of independent invariants of pure
		order $k$ equals $5$ for $k\geq 1$.
	\end{enumerate}
\end{theorem}

\subsection{Navier-Stokes invariants}

In this subsection we study the thermodynamic states that
admit a one-dimensional symmetry algebra generated by the
vector field $A$.

Considering the action of the thermodynamic vector field $A$
on the field of kinematic invariants and finding first
integrals of this action we get basis Navier-Stokes
differential invariants of the first order. 

Below we list basis Navier-Stokes invariants for the
different form of function $h(a)$.

\medskip
\textbf{1.} $h(a)=const$

When the thermodynamic state admits a one-dimensional
symmetry algebra generated by the vector field
\[
\xi_1 X_2 + \xi_2 X_3 + \xi_3 X_6 + \xi_4 X_7 =
\xi_1\dir{ \press}  + \xi_2 \dir{ \entr} + \xi_3
(t\,\dir{ t}+a\,\dir{a}-\press\,\dir{ \press} - \dens\,
\dir{\dens}) + \xi_4 (a\,\dir{a} + u\,\dir{ u} 
-{2  }\dens\,\dir{ \dens} +
{2  }\temp\,\dir{ \temp}),
\]
then the field of Navier-Stokes invariants is
generated by the first order differential invariants
\[
\entr + \frac{\xi_2}{\xi_3+2\xi_4}\ln\dens, \quad
u_a\dens^{-\frac{\xi_3}{\xi_3+2\xi_4}}, \quad
\dens_a\dens^{\frac{\xi_4}{\xi_3+2\xi_4}-2}, \quad
\frac{\dens^2(u_t+uu_a)}{\dens_au_a}, \quad
\frac{\dens\entr_a}{\dens_a}, \quad
\frac{\entr_t+u\entr_a}{u_a}
\]
and by the invariant differentiations 
\[
\dens^{-\frac{\xi_3}{\xi_3+2\xi_4}}\left( \totalDiff{t} +
u \totalDiff{a}\right)  ,\quad 
\dens^{-\frac{\xi_3+\xi_4}{\xi_3+2\xi_4}}\totalDiff{a} .
\]

\medskip
\textbf{2.}  $h(a)=\lambda a$, $\lambda\neq 0$

When the thermodynamic state admits a one-dimensional
symmetry algebra generated by the vector field
\begin{align*}
&\xi_1 X_2 + \xi_2 X_3 + \xi_3 X_6 + \xi_4 X_7 =
\xi_1\dir{ \press}  + \xi_2 \dir{ \entr} + \xi_3
(t\,\dir{ t}+2a\,\dir{a} + u\,\dir{ u}-\press\,\dir{ \press}
-3\dens\,\dir{ \dens} 
+ 2\temp\dir{ \temp}) +\\
&\phantom{jkhljkhljhjhkljhlkhlkjhljkhljhhjkgkjgkg}
\xi_4 \left(  t\,\dir{t}+
\left(\frac{\lambda g t^2}{2}+a\right)\dir{a} +
\lambda g t\,\dir{u}-\press\,\dir{\press} -
\dens\,\dir{\dens}\right) ,
\end{align*}
then the field of Navier-Stokes differential invariants is
generated by the differential invariants
\[
\entr + \frac{\xi_2}{3\xi_3+\xi_4}\ln\dens, \quad
u_a\dens^{-\frac{\xi_3+\xi_4}{3\xi_3+\xi_4}}, \quad
\dens_a\dens^{\frac{\xi_3}{3\xi_3+\xi_4}-2}, \quad
\frac{\dens^2(u_t+uu_a-\lambda g)}{\dens_au_a}, \quad
\frac{\dens\entr_a}{\dens_a}, \quad
\frac{\entr_t+u\entr_a}{u_a}
\]
of the first order and by the invariant differentiations 
\[
\dens^{-\frac{\xi_3+\xi_4}{3\xi_3+\xi_4}}\left( \totalDiff{t} + u \totalDiff{a}\right)  ,\quad 
\dens^{-\frac{2\xi_3+\xi_4}{3\xi_3+\xi_4}}\totalDiff{a} .
\]

\medskip
\textbf{3.}  $h(a)=\lambda a^2$, $\lambda\neq 0$

If the thermodynamic state admits a one-dimensional symmetry
algebra generated by the vector field
\[
\xi_1 X_2 + \xi_2 X_3 + \xi_3 X_6  = \xi_1\dir{ \press}  + \xi_2 \dir{ \entr} + \xi_3 ( a\, \dir{ a} +u\,\dir{ u}  
-2\dens\,\dir{ \dens} +
{2  }\temp\dir{ \temp} ) ,
\]
then the field of Navier-Stokes differential invariants is
generated by the first order differential invariants
\[
\entr + \frac{\xi_2}{2\xi_3}\ln\dens, \quad
u_a, \quad
\dens(u_t+uu_a-2\lambda ga)^2, \quad
\frac{\dens_a^2}{\dens^3}, \quad
\frac{\entr_a^2}{\dens}, \quad
\entr_t+u\entr_a
\]
and by the invariant differentiations 
\[
\totalDiff{t} + u \totalDiff{a},\quad 
\dens^{-\frac{1}{2}}\totalDiff{a} .
\]

\medskip
\textbf{4.}  $h(a)= \lambda_1a^{\lambda_2}$, $\lambda\neq 0, 1, 2$

If the thermodynamic state admits a one-dimensional symmetry
algebra generated by the vector field
\begin{equation*}
\xi_1 X_2 + \xi_2 X_3 + \xi_3 X_4 = \xi_1\dir{\press}+
\xi_2 \dir{ \entr} + \xi_3  \left( t\,\dir{ t} -
\frac{2a}{\lambda_2-2} \dir{ a} - \frac{\lambda_2 u }{\lambda_2-2}\dir{ u} - 
\press\,\dir{ \press}  +
\frac{\lambda_2 + 2  }{\lambda_2-2}\dens\,\dir{ \dens} -
\frac{2\lambda_2  }{\lambda_2-2}\temp\,\dir{ \temp} \right) ,
\end{equation*}
then the field of Navier-Stokes differential invariants is
generated by the first order differential invariants
\[
\entr + \frac{\xi_2(\lambda_2-2)}{2\xi_3}\ln a, \quad
 a^{-\lambda_2}u^2, \quad
au\dens,\quad
\frac{a u_t}{u^2}, \quad
\frac{au_a}{u}, \quad
a^2u \dens_a, \quad
\frac{a\entr_t}{u}, \quad a\entr_a
\]
and by the invariant differentiations 
\[
\dens^{\frac{\lambda_2-2}{\lambda_2+2}}\totalDiff{t}   ,\quad 
\dens^{\frac{-2}{\lambda_2+2}}\totalDiff{a} .
\]

\medskip
\textbf{5.}  $h(a)= \lambda_1e^{\lambda_2a}$

If the thermodynamic state admits a one-dimensional symmetry
algebra generated by the vector field
\begin{equation*}
\xi_1 X_2 + \xi_2 X_3 + \xi_3 X_4 = \xi_1\dir{ \press}+
\xi_2 \dir{\entr}+\xi_3\left(t\,\dir{t}-
\frac{2}{\lambda_2}\,\dir{a} - u\,\dir{u}-
\press\,\dir{ \press}  +
\dens\,\dir{\dens} -2\temp\dir{\temp}\right),
\end{equation*}
then the field of Navier-Stokes differential invariants is
generated by the first order differential invariants
\[
\entr + \frac{\lambda_2\xi_2}{2\xi_3} a, \quad
e^{-\lambda_2a}u^2, \quad
u\dens,\quad
\frac{ u_t}{u^2}, \quad
\frac{u_a}{u}, \quad
u \dens_a, \quad
\frac{\entr_t}{u}, \quad \entr_a
\]
and by the differentiations 
\[
\dens\,\totalDiff{t},\quad \totalDiff{a}.
\]

\medskip
\textbf{6.}  $h(a)=\ln a$

If the thermodynamic state admits a one-dimensional symmetry
algebra generated by the vector field
\begin{equation*}
\xi_1 X_2 + \xi_2 X_3 + \xi_3 X_4 = \xi_1\dir{ \press}  + \xi_2 \dir{ \entr} + \xi_3  \left( t\,\dir{ t}+ a  \,\dir{a}- \press\,\dir{ \press} -\dens\,\dir{\dens} \right),
\end{equation*}
then the field of Navier-Stokes differential invariants is
generated by the first order differential invariants
\[
\entr - \frac{\xi_2}{\xi_3} \ln a, \quad
u, \quad
a\dens,\quad
a u_t, \quad
au_a, \quad
a^2 \dens_a, \quad
a\entr_t, \quad a\entr_a
\]
and by the invariant differentiations 
\[
\dens^{-1}\totalDiff{t}   ,\quad 
\dens^{-1} \totalDiff{a} .
\]

\section*{Appendix}

The following table summarizes relations between the
function $h$ and the symmetry algebra of the system
\eqref{eq:E}, see Section \ref{sec:symmetries} for details.

\vspace*{10pt}
\begingroup
\setlength{\tabcolsep}{2pt}
\renewcommand{\arraystretch}{2.9}
\begin{tabular}{||b{0.26\linewidth}|b{0.66\linewidth}||}
	\hline
	$h(a)$ is arbitrary
	&$\begin{aligned} 
	&X_1 = \dir{ t}, \qquad\\
	&X_2 = \dir{\press }, \qquad \\
	&X_3 = \dir{\entr}
	\end{aligned} $ \\
	\hline
	$h(a)= const$ 
	&$\begin{aligned} 
	&X_4 = \dir{ a}, \qquad\\
	&X_5 = t\,\dir{ a}+\dir{ u}, \qquad \\
	&X_6 = t\,\dir{ t}+a\,\dir{a}-\press\,\dir{ \press} - \dens\,\dir{ \dens}, \\
	&X_{7} = a\,\dir{ a} + u\,\dir{ u} -2 \dens\,\dir{ \dens} + 2\temp\,\dir{ \temp}
	\end{aligned} $ \\
	\hline
	$h(a)= \lambda a$, $\lambda\neq 0 $ 
	& $\begin{aligned} 
	&X_4 = \dir{ a}, \qquad\\
	&X_5 = t\,\dir{ a}+\dir{ u}, \qquad \\
	&X_6 = t\,\dir{ t}+2a\,\dir{a} +  u\,\dir{ u} -\press\,\dir{ \press} -3 \dens\,\dir{ \dens} +2\temp\,\dir{ \temp}, \\
	&X_{7} =  t\,\dir{ t}+ (\frac{\lambda g t^2}{2}+a)\,\dir{ a} + \lambda g t\,\dir{ u}-\press\,\dir{ \press} - \dens\,\dir{ \dens}  
	\end{aligned} $ \\
	\hline
	$h(a)= \lambda a^2$, $\lambda\neq 0 $
	&$\begin{aligned} 
	&X_4 = \exp(\sqrt{2\lambda g}\,t)\,\dir{a} +  \sqrt{2\lambda g} \exp(\sqrt{2\lambda g}\,t)\,\dir{ u} , \qquad \\
	&X_5 =\exp(-\sqrt{2\lambda g}\,t)\,\dir{a} -  \sqrt{2\lambda g} \exp(-\sqrt{2\lambda g}\,t)\,\dir{ u}  \\
	&X_6 = a \dir{ a} +u\,\dir{ u}  -2 \dens\,\dir{ \dens} + 2\temp\,\dir{ \temp}, \qquad\\
	\end{aligned} $\\
	\hline
	$h(a)= \lambda_1a^{\lambda_2}$, $\lambda_2\neq 0,1,2 $
	&$\begin{aligned} 
	&X_4 = t\,\dir{ t} -\frac{2a}{\lambda_2-2} \dir{ a} - \frac{\lambda_2 u }{\lambda_2-2}\dir{ u} 
	- \press\,\dir{ \press}  +
	\frac{\lambda_2+2  }{\lambda_2-2}\dens\,\dir{ \dens} - \frac{2\lambda_2 \temp }{\lambda_2-2}\dir{ \temp}   \qquad
	\end{aligned}$\\
	\hline
	$h(a)= \lambda_1e^{\lambda_2a}$, $\lambda_2\neq 0 $
	&$\begin{aligned} 
	&X_4 = t\,\dir{ t}-\frac{2	}{\lambda_2}\,\dir{a} -  u\,\dir{ u} - \press\,\dir{ \press} +\dens\,\dir{\dens}-2\temp\,\dir{ \temp}
	\end{aligned}$\\
	\hline
	$h(a)= \ln{a}$ 
	&$\begin{aligned} 
	&X_4  = t\,\dir{ t}+ a  \,\dir{a}- \press\,\dir{ \press} -\dens\,\dir{\dens}
	\end{aligned}$\\
	\hline
\end{tabular}
\vspace{12pt}
\endgroup


\textbf{Acknowledgments.} The research was partially
supported by RFBR Grant No 18-29-10013.


\end{document}